\documentclass[aip,apl,reprint,a4paper,floatfix,groupedaddress,citeautoscript]{revtex4-2}
\usepackage{mathptmx}
\usepackage{helvet}
\usepackage[utf8]{inputenc} 
\usepackage{graphicx}
\usepackage{epstopdf}
\usepackage{amsmath}
\usepackage{xspace}
\usepackage{sidecap}
\usepackage{color,soul}
\usepackage[colorinlistoftodos,prependcaption,textsize=footnotesize,textwidth=1.4cm]{todonotes}
\usepackage[
,textwidth=17.5cm
,textheight=23.5cm
,verbose
,dvips
]{geometry}

\usepackage{hyperref}
\hypersetup{
  pdftitle={Sequential directional deposition of one-sided (In,Ga)N shells on GaN nanowires by molecular beam epitaxy},
	colorlinks,
	citecolor=blue,
	linkcolor=blue
	}

\newcommand{\celsius}{$^{\circ}$C\xspace}

\begin{document}

\title{Sequential directional deposition of one-sided (In,Ga)N shells on GaN nanowires by molecular beam epitaxy} 
\author{David van Treeck}
\thanks{D. van Treeck and J. L\"ahnemann contributed equally to the work}
\author{Jonas L\"ahnemann} 
\thanks{D. van Treeck and J. L\"ahnemann contributed equally to the work}
\email[Author to whom correspondence should be addressed: ]{laehnemann@pdi-berlin.de}
\author{Guanhui Gao}
\altaffiliation[Present address: ]{Department of Materials Science and NanoEngineering, Rice University, Houston, Texas, USA}
\author{Sergio Fern\'andez Garrido}
\altaffiliation[Present address: ]{Institute for Optoelectronic Systems and Microtechnology (ISOM) and Materials Science Department, Universidad Politécnica de Madrid, Avenida Complutense 30, 28040 Madrid, Spain}
\author{Oliver Brandt}
\author{Lutz Geelhaar}
\affiliation{Paul-Drude-Institut für Festkörperelektronik, Leibniz-Institut im Forschungsverbund Berlin e.V., Hausvogteiplatz 5--7, D-10117 Berlin, Germany}

%

\begin{abstract}
Capitalizing on the directed nature of the atomic fluxes in molecular beam epitaxy, we propose and demonstrate the sequential directional deposition of lateral (In,Ga)N shells on GaN nanowires. In this approach, a sub-monolayer thickness of each constituent atomic species, i.\,e. Ga, In, and N, is deposited subsequently from the same direction by rotating the sample and operating the shutters accordingly. Using multiple iterations of this process, we achieve the growth of homogeneous shells on a single side facet of the nanowires. For higher In content and thus lattice mismatch, we observe a strain-induced bending of the nanowire heterostructures. The incorporation of In and the resulting emission spectra are systematically investigated as a function of both the growth temperature and the In/Ga flux ratio.
\end{abstract}

\maketitle

\section{Introduction}

The three-dimensional geometry of nanowires (NWs) allows the growth of more complex heterostructures than possible for conventional planar thin films, which opens up new possibilities for device design.\cite{hyun_2013}  For example, in NWs, strain partitioning enables the accommodation of higher degrees of strain arising from lattice mismatch without plastic relaxation.\cite{Glas2015} Another benefit is that core-shell NW heterostructures offers the possibility for an increased volume per substrate area of the active region of semiconductor devices. This concept is attractive for light-emitting diodes and has been pursued, e.g., for the material system (In,Ga)N/GaN.\cite{Waag2011} Furthermore, the NW geometry can also facilitate more advanced heterostructures, such as one-sided shells. For classical III-V semiconductors, namely GaAs, GaP, InSb, and related alloys, the intentional deposition of a heterostructure on only one lateral surface of NWs has been demonstrated.\cite{Heigoldt_2009,Lewis2018a,Gagliano2018,Greenberg2019,McDermott2021} This selectivity is based on the directionality of fluxes in molecular beam epitaxy (MBE) and utilizes either tilted core NWs\cite{Heigoldt_2009} or ``digital alloy'' deposition,\cite{Lewis2018a} where the group-III elements of a ternary III-V shell are deposited in an alternating fashion. In contrast, for group-III nitrides, asymmetric NW shells have been reported only as a transitional phenomenon caused by inhomogeneous shell nucleation.\cite{Hetzl_2017}

For this specific material class of group-III nitrides, the formation of core-shell heterostructures in general is complicated by the peculiarities of the NW growth mechanisms in MBE.\cite{Foxon2009,vanTreeck2020,vanTreeck_arxiv_2023} Whereas conventional III-V NWs are typically grown by the vapor-liquid-solid mechanism, in which uniaxial growth is induced by a droplet,\cite{Guniat_2019} GaN and related NWs form spontaneously.\cite{FernandezGarrido_2015} Thus, the interplay between incorporation on the top and sidewall facets is particularly pronounced, and the enhancement of radial growth needed for shell formation is more difficult than for conventional III-V NWs. When samples are grown with continuous rotation and fluxes, the incidence of the constituent elemental fluxes from different directions together with the three-dimensional nature of the NWs leads to a situation which in planar growth is known as migration enhanced epitaxy:\cite{Horikoshi1986} on individual side facets, the different elements are deposited in an alternating sequence. However, in the case of GaN NWs, the permanent presence of N on the top facet causes this facet to constitute a pronounced sink for the metal atoms.\cite{vanTreeck2020} Thus, during the growth of (In,Ga)N shells the enhanced diffusion of In and Ga to the tip leads to the formation of large parasitic top segments.\cite{vanTreeck_arxiv_2023} This competition between the incorporation on the side and top facets of the NWs impedes the usability of such core-shell heterostructures for potential devices. To shift the balance of this competition toward incorporation on the side facets requires an improved deposition sequence that allows us to control which atomic species is present where and when.

In this paper, we present an innovative approach for the growth of (In,Ga)N shells on GaN nanowires that exploits the directed nature of the material beams in MBE: Only a single atomic species is deposited at a given time and the sample is rotated accordingly to enable sequential directional deposition (SDD) of either GaN or (In,Ga)N on only one side of GaN core NWs. This approach allows to reduce the diffusion of metal adatoms to the NW tip by avoiding  the permanent presence of active N species on this facet. This protocol enables the growth of uniform, one-sided shells on thin GaN NWs. From a more general perspective, our study of one-sided shell growth as a conceptionally simplified system helps to assess the potential of molecular beam epitaxy for the growth of core-shell structures in the nitride material system. In particular, in comparison to shell growth with continuous fluxes\cite{vanTreeck_arxiv_2023} the parasitic growth on top of the NWs is substantially diminished.



\section{Results and Discussion}
\subsection{Sequential directional deposition}

For these investigations, self-assembled GaN NWs on TiN grown at a V/III flux ratio of 2.5 and a substrate temperature of 780\,\celsius, as introduced in Refs.~\citenum{vanTreeck2018, vanTreeck2020}, are used as core NWs. These NW ensembles have a mean diameter and length of about 35 nm and 790 nm, respectively, at a density of about $10^9$\,cm$^{-2}$. This low area density facilitates studies of shell growth since shadowing by neighboring NWs is avoided.

\begin{figure}[t]
\centering
\includegraphics[width=\columnwidth]{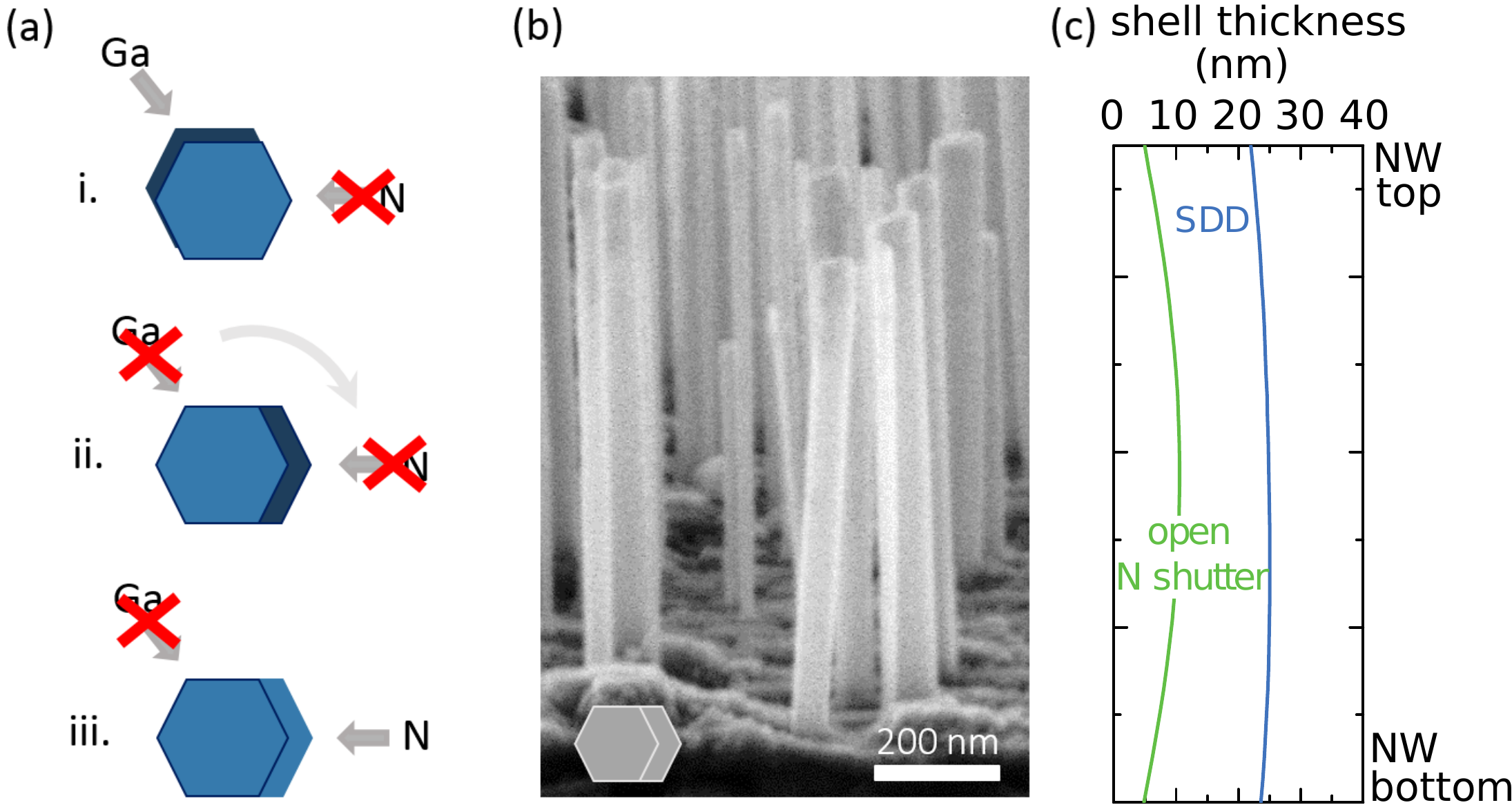}
\caption{\label{fig:binary} (a) Sketch of the SDD of a one-sided GaN shell. (b) Bird's eye SE micrograph of sample G, where one-sided GaN shells were grown on GaN core NWs using the SDD approach. (c) Modeled shell thickness of a one-sided GaN shell grown using the SDD approach (blue profile) and for constantly open N shutter (green profile) at otherwise similar modeling parameters.}
\end{figure}

To introduce the SDD approach, we start with binary GaN shells (sample G). Figure\;\ref{fig:binary}\;(a) outlines the SDD of a one-sided GaN shell on a GaN core NW. In the first step (i), Ga [typically $<1$ monolayer (ML)] is deposited on one specific side of the NW by opening the corresponding shutter for a few seconds. After closing the Ga shutter, the sample is rotated (ii) to align the very same side with the N source. In the final step (iii), the shutter of the N source is opened and GaN is grown. After closing the N shutter, the NW is rotated back to position (i) for the next deposition cycle.

Figure\;\ref{fig:binary}\;(b) shows a bird's eye secondary electron (SE) micrograph of the resulting sample G. The shell was grown at a substrate temperature of 490\,\celsius with a Ga-flux of 2.4\,nm/min and a N-flux of 12.4\,nm/min (direct fluxes in equivalent growth rate units for two-dimensional GaN layers). During each cycle, the NW side was exposed to the Ga beam for 4\,s---corresponding to the deposition of approximately 0.6\,ML of Ga on the exposed side facets and 0.8\,ML of Ga on the top facet/substrate, with the difference being caused by the different deposition angles in relation to the effusion cell. The wetted NW side was then rotated toward the N source and exposed to N for 25\,s. Both shutters were closed during the 4\,s long rotation steps. This cycle was repeated 130 times. Thus, a 20--25\,nm thick GaN shell was grown on the right side of the NWs as shown in the top view sketch in the inset to Fig.\;\ref{fig:binary}\;(b). The shell thickness was estimated by analyzing the NW diameter along the growth direction of the shell from SE micrographs. The one-sided shells are very homogeneous along the entire NW length, as seen in the micrograph of Fig.\;\ref{fig:binary}\;(b). This homogeneity is in contrast to the hourglass-shape obtained in  Ref.~\citenum{vanTreeck2020} using continuous Ga and N fluxes both for radially symmetric GaN shells obtained for continuous rotation and for one-sided shells grown without substrate rotation. However, note that the growth temperature in this study was reduced compared with Ref.~\citenum{vanTreeck2020}, in order to decrease adatom diffusivity, avoid Ga desorption and to subsequently enhance the incorporation of In in ternary (In,Ga)N shells.

To better understand the SDD, we modeled the shell thickness of a one-sided GaN shell on a core NW based on our work in Ref.~\citenum{vanTreeck2020}. Instead of the shell growth for \textit{continuous rotation} investigated there, we here simulated the different deposition and rotation sequences of the SDD mode. The model consists of a system of coupled one-dimensional diffusion equations for the different growth regions (top facet, sidewall facets, and substrate) that are solved numerically.  As input parameters, we used the growth conditions of sample G described above and modeling parameters corresponding to those from Ref.~\citenum{vanTreeck2020}. The resulting profile (blue) is shown in Fig.\;\ref{fig:binary}\;(c). For comparison, the green profile corresponds to a shell for constantly open N shutter with otherwise similar modeling parameters. In agreement with the experiment, the modeled profile for the SDD approach (blue) shows a rather uniform shell thickness along the NW with a slight barrel-shape.

During steps (i) and (ii), when the N shutter is still closed, the Ga adatom concentration on the side, the top facet and the substrate reaches an equilibrium state, where the surface is rather homogeneously wetted, explaining the homogeneous shell thickness. Note that due to the slightly higher adatom concentration on the top facet/substrate, compared to the side facet, there may be small diffusion fluxes onto the side facets. Only when the N-shutter is opened for step (iii), Ga adatoms diffuse from the NW side to the NW top and the substrate due to the higher incorporation rates in these regions (caused by the higher effective N flux related to the deposition geometry\cite{vanTreeck2020}). Due to these diffusion effects, the barrel shape is much more pronounced in case the N shutter is kept continuously open (green profile) and the increased Ga diffusion, away from the side facet, results in a much smaller mean shell thickness ($<10$\,nm).

\begin{figure}[t]
\centering
\includegraphics[width=\columnwidth]{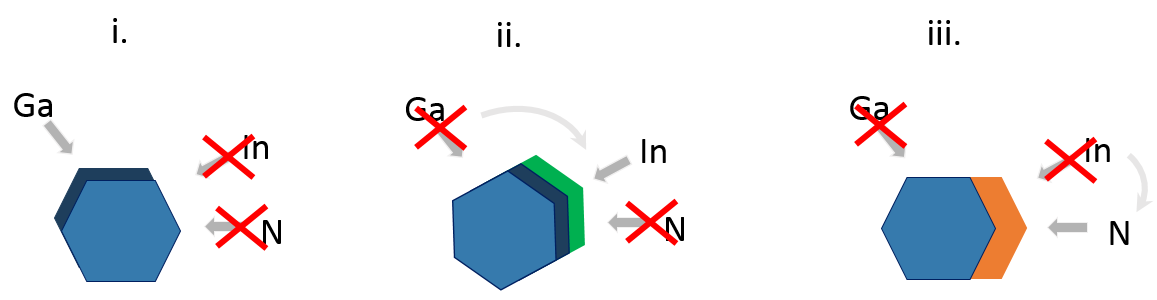}
\caption{\label{fig:sdd} Sketch of the SDD process of a one-sided (In,Ga)N shell with the source arrangement corresponding to our MBE system. (i) Ga deposition on the side of the NW. (ii) Rotation towards the In source and In deposition. (iii) Rotation towards the In source and opening the N shutter for (In,Ga)N growth. Note that all shutters are closed during the rotation phases.}
\end{figure}

\subsection{Systematic analysis of ternary alloy shells}

\begin{figure*}[t]
\centering
\includegraphics[width=0.99\textwidth]{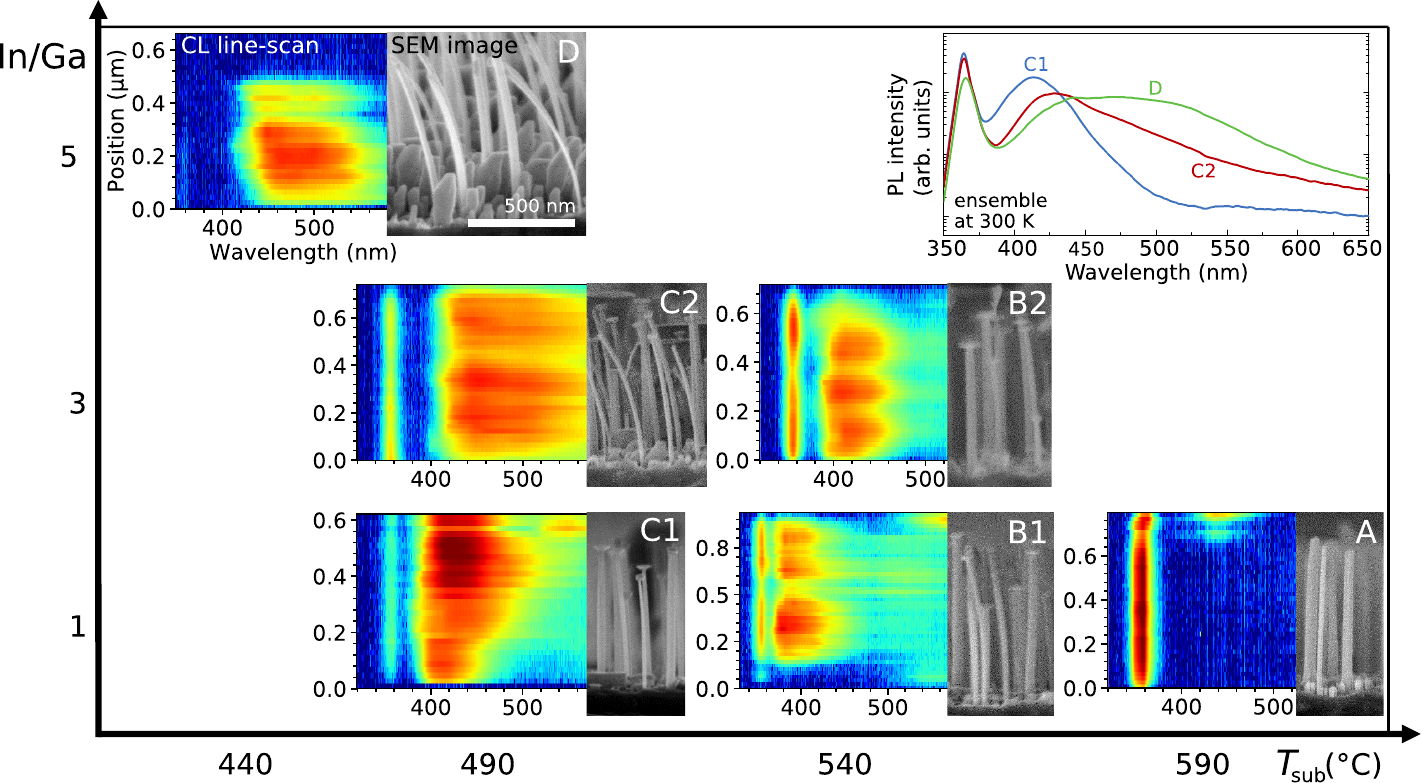}
\caption{\label{fig:map} SE micrographs of GaN NW ensembles with a one-sided (In,Ga)N shell and the respective CL line scans (measured at 10\,K) of single, dispersed NWs as a function of the growth temperature and the nominal In/Ga flux ratio. The CL intensities are color-coded in arbitrary units using a logarithmic scale. The top-right inset displays ensemble PL spectra of the samples C1, C2, and D acquired at room temperature. The growth conditions for the presented samples are summarized in Tab.\,\ref{tab:samp}.}
\end{figure*}

Turning to ternary alloy shells, Fig.\;\ref{fig:sdd} outlines the SDD process for a one-sided (In,Ga)N shell. Compared to the binary shell, an intermediate step is added after the Ga deposition, where the growth side faces the In cell and In is deposited. The total amount of deposited metal was kept constant at 0.3\,ML (0.4\,ML) on the side facet (top facet) per deposition cycle. Finally, the facets were exposed to an N flux of 16.2\,nm/min for approximately 25\,s seconds. 

A series of six samples with different In/Ga flux ratios (between 1 and 5) and at different growth temperatures (between 440 and 590\,\celsius) were grown, as summarized in Tab.~\ref{tab:samp}. Only for sample D, with an In/Ga flux ratio of 5, the metal flux had to be increased to 0.45\,ML (0.6\,ML) for the side facet (top facet) to maintain a reasonable calibration of the Ga flux and shutter response times. By adjusting the number of SDD cycles for each sample, we aimed to maintain a shell thickness of 10--15\,nm.

The morphology and emission properties of these GaN NWs with a one-sided (In,Ga)N shell (samples A--D), plotted as a function of the growth temperature and the nominal In/Ga flux ratio, are shown in Fig.\;\ref{fig:map}. The presented line scans are representative choices among 10--20 line scans on dispersed NWs performed for each sample. To resolve the emission along the NW axis, the NWs were dispersed on an Au-coated Si substrate and cathodoluminescence (CL) spectral line scans were measured using a Gatan MonoCL4 system mounted on a Zeiss Ultra55 field-emission scanning electron microscope (SEM) operated at 5\,kV acceleration voltage with a probe current of 1--1.5\,nA. The data was analyzed and plotted using the python packages HyperSpy\cite{HyperSpy} and LumiSpy.\cite{LumiSpy} In Fig.\;\ref{fig:map}, representative CL line scans collected at a sample temperature of 10~K on individual NWs are plotted alongside bird's eye scanning electron (SE) micrographs of the corresponding NW ensemble. The SE micrographs of the single NWs corresponding to each of the line scans are presented in the Supplementary Material. With respect to the CL measurements, in addition to the broad emission band of the (In,Ga)N shell at longer wavelengths, all samples also exhibit a notable luminescence at around 357\,nm (3.47\,eV), which can be attributed to the near band edge emission of the GaN cores. The contribution of this band varies between individual line scans even from the same sample, which might just be a geometric effect of the CL excitation depending on where the shell is oriented on the dispersed NWs. Note that the streaky features in the CL linescans are an experimental artefact due to the drift correction employed during the measurements.

In terms of morphology, the growth map reveals that with decreasing growth temperature from 540\,\celsius to 440\,\celsius and increasing In/Ga flux ratio from 1 to 5, the NWs bend more and more to one side, while the CL emission of the (In,Ga)N shell shifts from around 380\,nm (3.26\,eV) for sample B1 to about 480\,nm (2.58\,eV) for sample D. Only for sample A, grown at the highest temperature, no (In,Ga)N shell emission can be detected. Note that the one-sided (In,Ga)N shells show luminescence even without passivation by an outer GaN shell, indicating that the localization of carriers in the ternary alloy prevents them from recombining non-radiatively at the surface, as observed previously.\cite{vanTreeck_arxiv_2023} However, such carrier localization and variations in its effectiveness can lead to the observed variations in (In,Ga)N emission intensity along the length of single NWs, most pronounced for the line scan displayed for sample B2. Note that the apparent changes in the full width at half maximum of the emission band, correlating with these intensity variations, result from the logscale intensity colorscale chosen for the plots.
Additional experiments presented in the Supplementary Material show that a conventional co-deposition of all constituents without rotation also leads to one-sided shells. However, this approach results in very rough shells with an inhomogeneous composition. Furthermore, for SDD  the deposition sequence (Ga-In-N) is important to incorporate In, as also discussed in the Supplementary Material.

The bending of the NWs visible in the SE micrographs is a consequence of the larger lattice constant of (In,Ga)N with respect to GaN. Hence, once an (In,Ga)N shell is grown on one side of the GaN NWs (right side in the micrographs of Fig.\,\ref{fig:map}), they bend to the opposite (left) side. Thus, the bending of the NWs confirms the presence of a one-sided (In,Ga)N shell, with a larger In content and thus lattice mismatch resulting in a more pronounced bending of the NWs. For sample A, the straight NWs indicate no or only little In incorporation into the shell, consistent with the absence of a pronounced (In,Ga)N emission. Correspondingly, the red-shift of the shell luminescence with decreasing growth temperature can be attributed to an increase in In incorporation resulting from a reduced desorption of In and a lower dissociation rate of InN.\cite{Duff2014, Siekacz2011, Lang2012a, Ambacher1998a} 
Note that the bending of a NW results in a non-uniform strain distribution in the core-shell structure, which includes a gradient in the $\epsilon_{zz}$ component of the strain tensor across the NW diameter, affecting the band structure\cite{Lewis2018a} and hence the emission wavelength of the (In,Ga)N shell.

The increasing In/Ga flux ratio from samples B1 to B2 (and C1 to C2) further increases the In incorporation, corresponding to an additional red-shift and broadening of the shell emission. This red-shift and broadening is also visible when analyzing the ensemble photoluminescence (PL) spectra of the samples C1 and C2 shown in the top-right inset of Fig.\,\ref{fig:map}. The spectra were acquired at 300 K using a Horiba Jobin Yvon (Labram HR 800 UV) \textmu-PL setup with excitation from a HeCd laser (325\,nm) with a spot diameter of about 3\,\textmu{}m, thus probing about 70 NWs. The PL spectrum of sample D, grown at the lowest growth temperature and highest In/Ga flux ratio, shows an even broader emission band with rather constant emission intensities between 440\,nm (2.82\,eV) and 510\,nm (2.43\,eV).

For planar (In,Ga)N layers,\cite{Pereira_2001,Schley2007} the observed red-shift from 380\,nm (3.26\,eV) to 480\,nm (2.58\,eV) between samples B1 and D would correspond to a change of the In content from 6\% to 22\%. For our samples, this composition can only be a rough estimate, since the different strain state of the one-sided (In,Ga)N shells is not taken into account. With respect to the nominal In/Ga flux ratios used during the growth, only a small fraction of the In is incorporated. This discrepancy is not surprising, since even for sample D, In desorption was detected by line-of-sight quadruple mass spectrometry. The desorption being another loss channel in addition to the discussed diffusion of In from the sidewalls to the top facet of the NW and the substrate.

\begin{table}[b]
\begin{tabular}{cccccc}
\hline
Sample & $T_\mathrm{sub}$ & In/Ga & In per & Ga per & \# cycles\\[-.7mm]
 &  ($^\circ$C) & flux ratio &  cycle (ML) & cycle (ML) & \\
\hline
A & 590 & 1 & 0.2 & 0.2 & 250\\
B1 & 540 & 1 & 0.2 & 0.2 & 250\\
B2 & 540 & 3 & 0.3 & 0.1 & 350\\
C1 & 490 & 1 & 0.2 & 0.2 & 250\\
C2 & 490 & 3 & 0.3 & 0.1 & 350\\
D & 440 & 5 & 0.5 & 0.1 & 250\\
\hline
\end{tabular}
\caption{\label{tab:samp} Summary of the growth conditions for the investigated samples, including the growth temperature, the In/Ga flux ratio, the amount of deposited In and Ga per deposition cycle, as well as the total number of cycles carried out for the respective sample.}
\end{table}

For radially symmetric (In,Ga)N shells,\cite{vanTreeck2018} such a redshift of the shell emission with increasing In/Ga flux ratio has not been observed. The higher In incorporation for one-sided shells can be explained by a reduced In diffusion away from the side facet during the SDD process due to the sequential and directional N exposure. However, a broadening of the PL spectrum towards higher In/Ga flux ratios was also observed for the radially symmetric shells.\cite{vanTreeck_arxiv_2023} For the one-sided (In,Ga)N shells, this broadening may be attributed to the increasing variation of the emission wavelength from NW to NW observed between CL line scans, but also the emission band in individual NWs is significantly broadened with increasing In content. Note that different strain states along the axis of the one-sided (In,Ga)N shells due to the bending of the NW may also contribute to the emission broadening.\cite{Lewis2018a}

In Ref.~\citenum{vanTreeck_arxiv_2023}, we observed significant axial growth on the top facet for NWs with radially symmetric shells grown with continuous fluxes, which would complicate the processing of devices out of such NW heterostructures. This parasitic growth is notably reduced for our one-sided (In,Ga)N shells. The reduced volume of the top segment can be attributed to the smaller amount of material that is deposited, but also to the aforementioned reduction of adatom diffusion towards the NW top for the SDD approach. Since the CL analysis of the different samples focused on the emission properties of the shells, the spectral region where the top segments might emit was not covered. Only for samples B1 and C1, the onset of the top segment emission is visible at about 560\,nm (2.21\,eV), while for sample A, the NW tip is the only part showing (In,Ga)N emission.

\begin{figure}[t]
\centering
\includegraphics[width=\columnwidth]{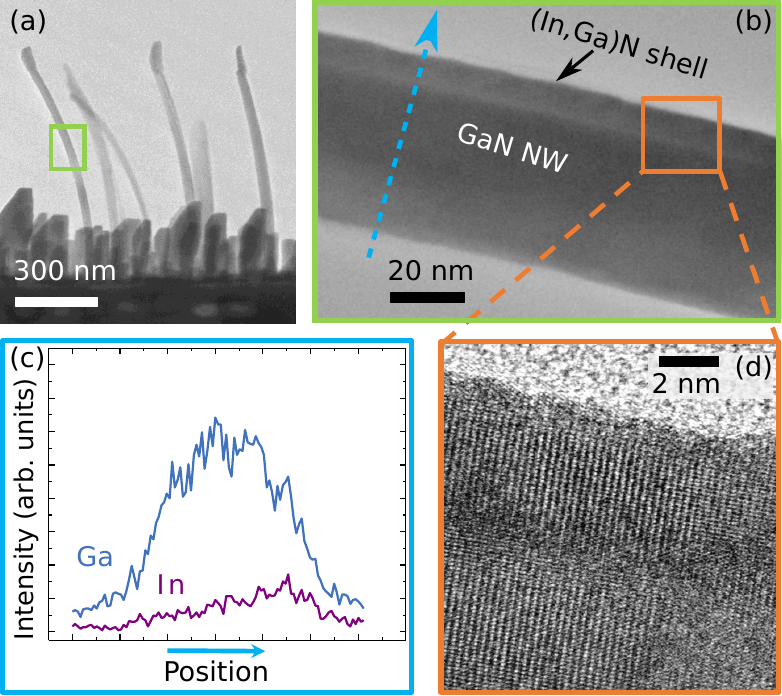}
\caption{\label{fig:tem} (a) Cross-sectional, bright-field STEM micrograph of sample D and (b) zoom into the mid segment of a representative NW. (c) STEM-EDX intensity profiles of the In and Ga signal across the NW diameter (blue arrow) of the NW presented in (b). (d) Exemplary high-resolution STEM micrograph of the interface between the (In,Ga)N shell and the GaN core NW along the [$10\bar{1}0$] zone axis.}
\end{figure}

To independently verify the presence of one-sided (In,Ga)N shells, Fig.\,\ref{fig:tem} shows the results of an analysis of sample D by scanning transmission electron microscopy (STEM) and STEM-based energy-dispersive X-ray spectroscopy (EDX) in a JEOL JEM-3010 microscope. The STEM sample was prepared using a classic polishing procedure where the NW ensemble was embedded in epoxy glue. The bending of the NWs and the parasitic underlayer are highlighted in Fig.\,\ref{fig:tem}(a). As already observed in SE micrographs, both the bending radius and the shape of the top segments can vary from NW to NW. Figure\,\ref{fig:tem}(b) shows an enlarged bright-field STEM micrograph of the middle segment of a representative NW. The approximately 10-nm-thick shell is clearly visible on only one side of the NW, while the opposite side shows no shell growth. A STEM-EDX line scan across the NW diameter, displayed in Fig.\,\ref{fig:tem}(c), shows a clear increase in the In signal for the position of the shell, proving that the shell does indeed contain a significant amount of In, even if a quantification is not possible. Note that the low intensity of the In-L lines results in a rather noisy profile with a notable contribution from the background. Finally, Fig.\,\ref{fig:tem}(d) shows a representative high-resolution micrograph of the interface between the (In,Ga)N shell and the GaN core NW. Analysis of such high resolution images along the NW showed that there is no evidence of plastic relaxation or dislocations in the shells. Apparently, most of the strain resulting from the different lattice constants of GaN and (In,Ga)N is released by the bending of the NW.

\section{Conclusions}

We have presented a sequential directional deposition method for GaN and (In,Ga)N using MBE to grow one-sided shells on GaN NW cores, while minimizing the parasitic growth on the top facet of the NW compared to shell growth employing continuous fluxes. Using a dedicated sequence of shutter openings and rotations, the individual metal fluxes and subsequently the nitrogen flux are directed towards the same facet. For the growth of GaN shells, the SDD mode allows not only the deposition of single shells on selected lateral sides of a NW, but also enables the growth of a homogeneous GaN shell along the entire NW length. For ternary, one-sided (In,Ga)N shells, the emission wavelength can be shifted in a controlled manner from the UV to the green range (510\,nm) by lowering the growth temperature (down to 440\,\celsius) and increasing the In/Ga flux ratio (up to 5), while the shell thickness remains homogeneous along the NW length. Using thin core NWs (average diameter 35\,nm), the one-sided (In,Ga)N shells can even induce a bending of the NWs. 

Our new approach to the growth of ternary (In,Ga)N shells can be exploited for device fabrication in two different ways. First, by applying the same sequential directional deposition sequence to all six NW sidewalls, radially symmetric homogeneous core-shell quantum wells can be grown with minimal parasitic axial growth. This benefit compared to continuous deposition simplifies the subsequent processing into devices such as light-emitting diodes. Second, new opportunities for device design are opened up. More specifically, one-sided shell quantum wells with different emission wavelengths can be grown on different sidewall facets of one and the same NW, thus potentially enabling multi-color light-emitting diodes within a single NW.

\section*{Supplementary Material}

See Supplementary Material for corresponding SE micrographs for the single NW CL line scans displayed in Fig.~\ref{fig:map}, as well as an investigation of one-sided shell growth with continuous fluxes (co-deposition) and samples grown with an inverted metal deposition sequence (In-Ga-N).

\acknowledgments

The authors would like to thank Andrea Ardenghi for a critical reading of the manuscript. The authors also thank Katrin Morgenroth, Carsten Stemmler and Michael Höricke for the maintenance of the MBE system, as well as Anne-Kathrin Bluhm for the SE micrographs. Sergio Fernández-Garrido acknowledges the partial financial support from the Spanish program Ramón y Cajal (co-financed by the European Social Fund) under Grant No. RYC-2016-19509 from the Ministerio de Ciencia, Innovación y Universidades.

\section*{Author Declarations}
\subsection*{Conflict of Interest}

The authors have no conflicts to disclose.

\section*{Data Availability Statement}

The data that support the findings of this study are available
within the article and its Supplementary Material. The raw
data may be obtained from the corresponding author upon
reasonable request.

\section*{References}
\bibliography{vanTreeck_apl}

\end{document}


\title{Supplementary Material: Sequential directional deposition of one-sided (In,Ga)N shells on GaN nanowires by molecular beam epitaxy} 
\author{David van Treeck}
\thanks{D. van Treeck and J. L\"ahnemann contributed equally to the work}
\author{Jonas L\"ahnemann} 
\thanks{D. van Treeck and J. L\"ahnemann contributed equally to the work}
\email[Author to whom correspondence should be addressed: ]{laehnemann@pdi-berlin.de}
\author{Guanhui Gao}
\altaffiliation[Present address: ]{Department of Materials Science and NanoEngineering, Rice University, Houston, Texas, USA}
\author{Sergio Fern\'andez Garrido}
\altaffiliation[Present address: ]{Institute for Optoelectronic Systems and Microtechnology (ISOM) and Materials Science Department, Universidad Politécnica de Madrid, Avenida Complutense 30, 28040 Madrid, Spain}
\author{Oliver Brandt}
\author{Lutz Geelhaar}
\affiliation{Paul-Drude-Institut für Festkörperelektronik, Leibniz-Institut im Forschungsverbund Berlin e.V., Hausvogteiplatz 5--7, D-10117 Berlin, Germany}

%

\maketitle

\section{Single nanowire SE micrographs}

\begin{figure}[t]
\includegraphics[width=\columnwidth]{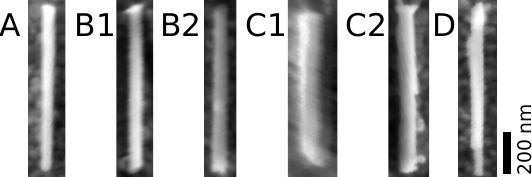}
\caption{\label{fig:se} Single nanowire SE micrographs corresponding to the NWs for which CL line scans are displayed in Fig.\,3 of the main manuscript.}
\end{figure}

Secondary electron (SE) micrographs of the exemplary single nanowires (NWs) from each sample for which hyperspectral cathodoluminescence (CL) line scans are displayed in Fig.\,3 of the main manuscript are provided in Fig.\,\ref{fig:se}. The diameter of the NWs ranges from 55~nm for the NWs from samples A and D to 90~nm for the NW from sample C1. These NW diameters are above the expected average diameters of 45--50~nm for the NW ensembles. Due to experimental constraints, fairly straight and thus especially for the NW ensembles with higher In content rather thicker NWs were chosen for the line scans. Also, as the thicker NWs have a larger (In,Ga)N volume, they generally exhibit a stronger luminescence signal. Nevertheless, the emission wavelengths are representative for the NW ensembles as can be seen from a comparison with the room-temperature ensemble photoluminescence (PL) measurements depicted in the inset of Fig.~3 of the main manuscript.

\section{Growth with continuous fluxes}

\begin{figure}[t]
\includegraphics[width=\columnwidth]{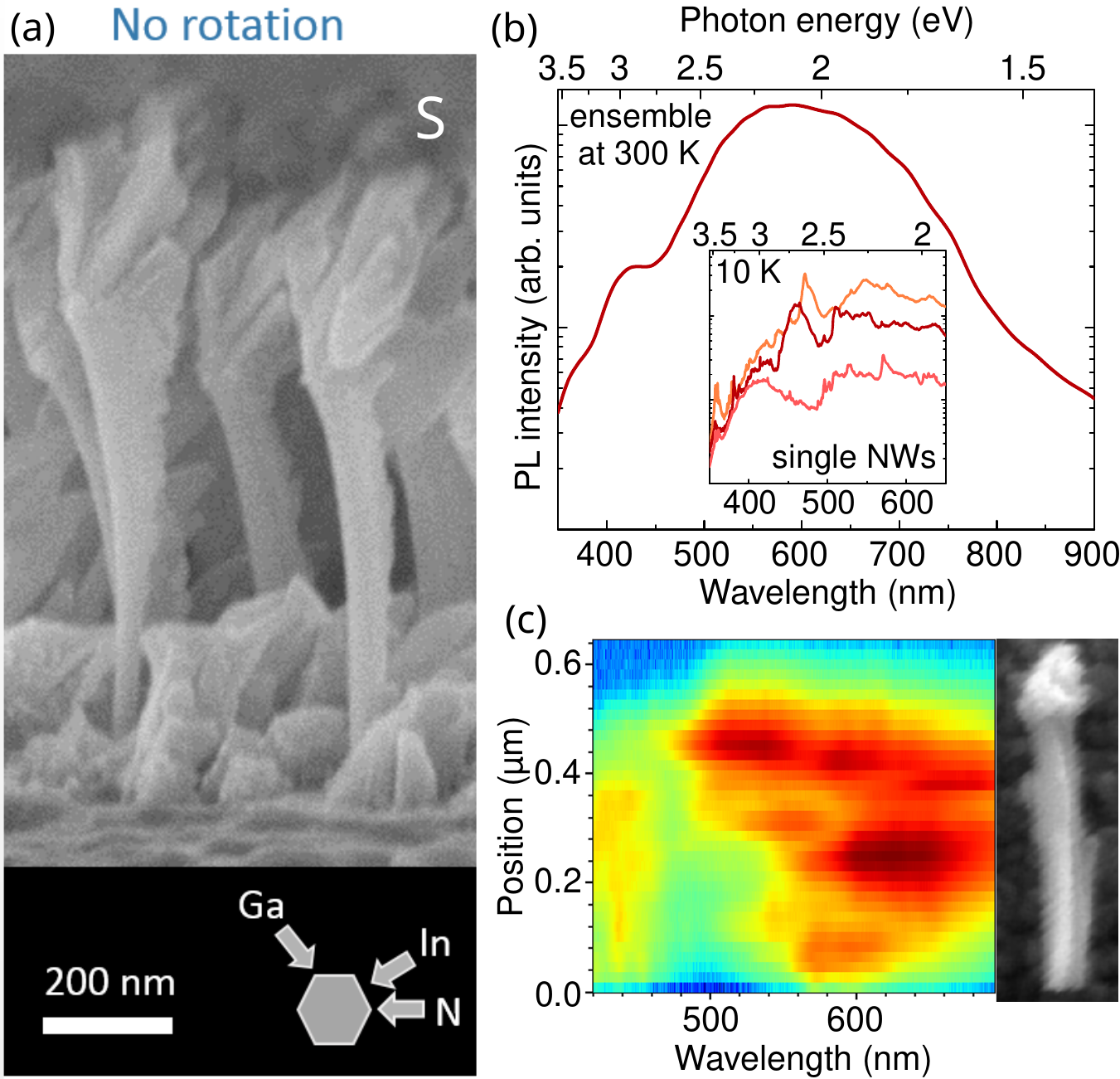}
\caption{\label{fig:norot} Cross-sectional SE micrograph of sample S, which was grown without substrate rotation with all atomic species impinging simultaneously. The sketch at the bottom depicts the source arrangement of the MBE chamber. (b) room-temperature PL spectrum of the NW ensemble measured at 300\,K. The inset depicts low-temperature PL measurements (at 10\,K) of exemplary single NWs. (c) Low-temperature CL line scan of a representative single NW.}
\end{figure}

To emphasize the advantage of sequential directional deposition (SDD) for obtaining homogeneous one-sided shells, sample S shown in Fig.\,\ref{fig:norot}(a) was grown without substrate rotation while co-depositing all atomic species with equal In and Ga fluxes of 0.1~ML/s and a V/III ratio of 5 at a substrate temperature $T_\mathrm{sub}=490$~\celsius. 
The NWs show pronounced strain-induced bending, but instead of a homogeneous shell, large fan-like structures have formed on one side of the NWs. The fan structures can be up to 200~nm thick, are rather irregular along the NW length, and also differ significantly from NW to NW. In addition, the upper segments are very pronounced and tilted to the right-hand side.

The NW morphology shows that the radial growth of (In,Ga)N  takes place predominantly on the N-exposed side of the NWs (right-hand side), similar to the case without rotation for GaN shells in Ref.~\citenum{vanTreeck2020}. Since the N-exposed side is shadowed from the direct Ga beam, the Ga adatoms contributing to the sidewall growth must have diffused to the growth front. 

Figure\,\ref{fig:norot}(b) shows a photoluminescence (PL) measurement of the NW ensemble of sample~S at 300\,K. The spectrum is extremely broad with a main emission band around 590\,nm (2.1\,eV), a FWHM of 200\,nm (690\,meV) and a pronounced shoulder at about 420\,nm (2.95\,eV). The inset shows low-temperature PL measurements at 10\,K, where each spectrum corresponds to the PL of one or few NWs (the number of NWs was not clearly resolvable). All spectra show a large number of narrow emission lines at different wavelengths throughout the measurement range from 350\,nm (3.54\,eV) to 650\,nm (1.91\,eV).  
A representative CL line scan is shown in Fig.\,\ref{fig:norot}(c). In contrast to the SDD grown samples, the emission is significantly redshifted, but very inhomogeneous along the NW length. A number of localized emission centers at different wavelengths from 530\,nm (2.34\,eV) to 670\,nm (1.85\,eV) are seen. In addition, a low-intensity emission band is visible at about 430\,nm (2.88\,eV). Although a high In content is achieved, the composition varies greatly leading to charge carrier localization.\cite{Albert2014,Coulon2017,Kapoor2018}

The higher In incorporation on the NW sidewall may have several reasons. Since the side facet, where growth takes place, is constantly exposed to both In and N, there are always N-rich conditions at the growth front, which in combination with a permanent In supply favor high InN mole fractions in (In,Ga)N.\cite{Siekacz2011,Lang2012a} 
Moreover, it is likely that with increasing (In,Ga)N thickness on the \textit{M}-plane, local strain variations caused by fluctuations in the In content may induce the formation of other facets.\cite{Coulon2017} The inclined nature and coarse morphology of the (In,Ga)N fan structures suggest the formation of semi-polar facets along which further growth occurs. 
On semi-polar facets, the efficiency of In incorporation is said to be higher than on the \textit{M}-plane\cite{Wernicke2012k,Coulon2017}.

The emission band around 430\,nm (2.88\,eV), which is clearly visible in both the RT-PL and the LT-CL line scan, also seems to originate from the NW sidewall and could be attributed to an initial layer with low In content.

\begin{figure}
\includegraphics[width=\columnwidth]{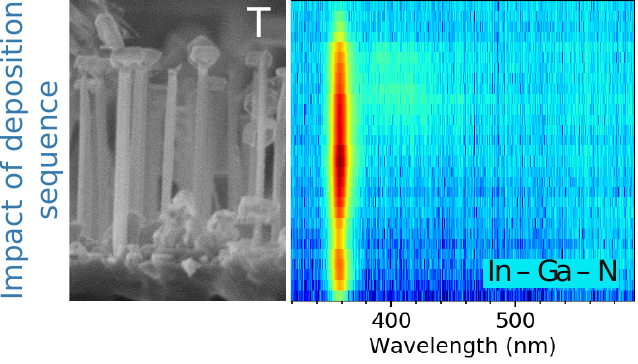}
\caption{\label{fig:var} SE micrograph of the NW ensemble and CL line scan (measured at 10\,K) of a representative, dispersed NW for sample T, where the deposition sequence of the different atomic species was changed from Ga first to In first. The CL intensities are color-coded in arbitrary units using a logarithmic scale.}
\end{figure}

\section{Role of the deposition sequence}

In the following, we will examine the influence of the deposition sequence on the final In content of the shells.

For sample T, shown in Fig.~\ref{fig:var}(b), the deposition sequence was changed to start with In (In-Ga-N) instead of with Ga (Ga-In-N). Otherwise, the growth conditions were the same as for sample C1. The emission at 430\,nm is absent in the CL line scan despite the formation of a one-sided shell (deduced from the increase in NW diameter). Thus, the change in the rotation sequence prevented the incorporation of In into the shell. Consistent with this interpretation, a dramatically increased desorption of In after opening the Ga shutter was detected by line-of-sight quadrupole-mass spectrometry (QMS).

Since Ga is deposited after In, it is likely that the adsorbed In atoms are displaced from the energetically favorable sites on the surface by Ga adatoms due to the more favorable bonding configuration. Indeed, strong surface segregation of In during the growth of In is universally observed experimentally and is understood to be a consequence of both the strain introduced by In and the weaker bond strength of In-N compared to Ga-N.\cite{Chen2000,Waltereit2002,Duff2014} The exchange process between In and Ga increases the probability of subsequent In desorption from the surface, as observed by QMS, resulting in a very low or no In incorporation at all. In the original scenario, where Ga is deposited first (Ga-In-N sequence), this exchange process is unlikely to occur because most Ga adatoms have already found a stable configuration when In is deposited. Thus, the probability for In adatoms to form a stable bond on the surface and to be incorporated into the lattice once N is supplied is significantly higher in this scenario.

Overall, the growth conditions and the sequence of the different shells have to be carefully planned and optimized to obtain one-sided (In,Ga)N shell growth.

\vspace{7mm}

\section*{References}
\bibliography{vanTreeck_apl}